\documentclass[runningheads]{llncs}

\usepackage{graphicx}
\usepackage{booktabs}
\usepackage{multirow}
\usepackage{subfig}
\usepackage{cite}
\usepackage{hyperref}

\newlength\lengtha 
\newlength\lengthb 

\begin{document}

\title{Bayesian Neural Networks for Uncertainty Estimation of Imaging Biomarkers}

\author{J. Senapati$^a$ \and A. Guha Roy$^a$ \and S. P\"olsterl$^a$ \and D. Gutmann$^b$ \and S. Gatidis$^b$ \and C.~Schlett$^c$ \and A. Peters$^d$ \and F. Bamberg$^c$ \and C. Wachinger$^a$}


\authorrunning{Senapati et al.}

\institute{
$^a$Artificial Intelligence in Medical Imaging (AI-Med), KJP, LMU M\"unchen, Germany \\
$^b$Department of Diagnostic and Interventional Rad., University of T\"ubingen, Germany \\
$^c$Department of Diagnostic and Interventional Rad., University Freiburg, Germany\\
$^d$Institute of Epidemiology, Helmholtz Zentrum M\"unchen, Germany
}

\maketitle            

\begin{abstract}

Image segmentation enables to extract quantitative measures from scans that can serve as imaging biomarkers for diseases. 
However,  segmentation quality can vary substantially across scans, and therefore yield unfaithful estimates in the follow-up statistical analysis of biomarkers. 
The core problem is that segmentation and biomarker analysis are performed independently.
We propose to propagate segmentation uncertainty to the statistical analysis to account for variations in segmentation confidence. 
To this end, we evaluate four Bayesian neural networks to sample from the posterior distribution and estimate the uncertainty. 
We then assign confidence measures to the biomarker and propose statistical models for its integration in group analysis and disease classification. 
Our results for segmenting the liver in patients with diabetes mellitus clearly demonstrate the improvement of integrating biomarker uncertainty in the statistical inference. 

\end{abstract}

\section{Introduction}

Imaging biomarkers play a crucial role in tracking disease progression, in supporting an automated prediction of diagnosis, and in providing novel insights in the pathophysiology of diseases~\cite{wachinger2016whole,wachinger2016domain,becker2018gaussian}. 
A prerequisite for many image-based markers is image segmentation, which provides access to morphological features like volume, thickness and shape information~\cite{gutierrez2018deep,gutierrez2019learning,wachinger2015brainprint}. 
Despite a boost in segmentation accuracy by deep learning~\cite{minaee2020image}, automated segmentations are not perfect and their quality can vary substantially across scans.  
As a consequence, segmentation errors propagate to errors in the derived biomarker. 
To reduce the impact of erroneous segmentations in follow-up analyses and to infer faithful estimates, a manual quality control is advised to identify segmentations of sufficient quality. 
However, the manual quality assessment is subject to intra- and inter-rater variability and time consuming, particularly for large datasets.

Fortunately, Bayesian neural networks for image segmentation~\cite{kendall2017uncertainties, kohl2018probabilistic,kohl2019hierarchical}
have been developed that do not only provide the mode (i.e., the most likely segmentation) but also the posterior distribution of the segmentation.
Monte Carlo (MC) dropout~\cite{kendall2017uncertainties,gal2016dropout} or the probabilistic U-Net~\cite{kohl2018probabilistic,kohl2019hierarchical}  enable to sample multiple possible segmentations instead of only a single segmentation. 
Typically, a voxel-wise uncertainty is then computed and displayed to the user to detect regions with lower segmentation confidence
In contrast, we want to use the segmentation uncertainty to derive a biomarker uncertainty. 
With such a measure, we could directly determine the scans from which biomarkers have been  extracted reliably without the need for a manual quality control. 
However, the integration of the segmentation uncertainty into follow-up analyses of extracted biomarkers, such as group analyses or disease classification, has not yet been well studied. 
In addition, it is not clear which Bayesian segmentation method is best suited for inferring the uncertainty of the biomarker, capturing different aspects of aleatoric and epistemic uncertainty. 

To address these issues, we present statistical models that integrate  segmentation confidence measures in the parameter inference of the biomarker. 
Further, we compare four state-of-the-art Bayesian neural networks for computing the segmentation and confidence measures. 
We perform experiments for the segmentation of the liver in abdominal magnetic resonance imaging (MRI) scans in subjects with diabetes mellitus. 
Our results demonstrate that the integration of biomarker uncertainty yields estimates that are closer to the manual reference and higher classification accuracy. 

\textbf{Related Work} 
Several approaches have been proposed to compute uncertainty for the segmentation of medical images~\cite{nair2020exploring,jungo2019assessing,sedai2019uncertainty,yu2019uncertainty,baumgartner2019phiseg,roy2019bayesian,eaton2018towards,hu2019supervised}.
We have previously used MC dropout to compute the uncertainty in whole-brain segmentation~\cite{roy2018inherent,roy2019bayesian}.
Nair et al.\cite{nair2020exploring} provide four different voxel-based uncertainty measures based on MC dropout. 
The reliability of uncertainty estimations for image segmentation has been evaluated in~\cite{jungo2019assessing}.
Eaton et al.~\cite{eaton2018towards} presented uncertainty for calibrating confidence boundary for robust predictions of tumor segmentations.
Hu et al. \cite{hu2019supervised} used calibrated inter-grader variability as a target for training a CNN model to predict aleatoric uncertainty. 
Sedai et al. \cite{sedai2019uncertainty} used uncertainty guided domain alignment to train a model for retinal and choroidal layers segmentation in OCT images. 
In \cite{yu2019uncertainty}, Yu et al. incorporate uncertainty in a teacher CNN to guide a student CNN in a semi-supervised segmentation setup. The architecture for multi-scale ambiguity detection and uncertainty quantification proposed by Baumgartner et al.~\cite{baumgartner2019phiseg} is similar to the hierarchical model in~\cite{kohl2019hierarchical}, which we also use in this work.
Uncertainty-driven bootstrapping for data selection and model training was proposed in~\cite{ghesu2019quantifying}.
In contrast to prior work, we focus on propagating segmentation to the statistical analysis of imaging biomarkers. 

\section{Methods}

\subsection{Bayesian Neural Networks for Image Segmentation}
Essential for our approach of estimating the biomarker uncertainty are Bayesian segmentation networks that enable to sample from the predictive posterior distribution. 
Several strategies to perform variational inference within fully convolutional neural networks (F-CNNs) have been proposed in the literature, where we describe four commonly used and promising approaches in the following. 

\begin{figure}[t]
\includegraphics[width=\textwidth]{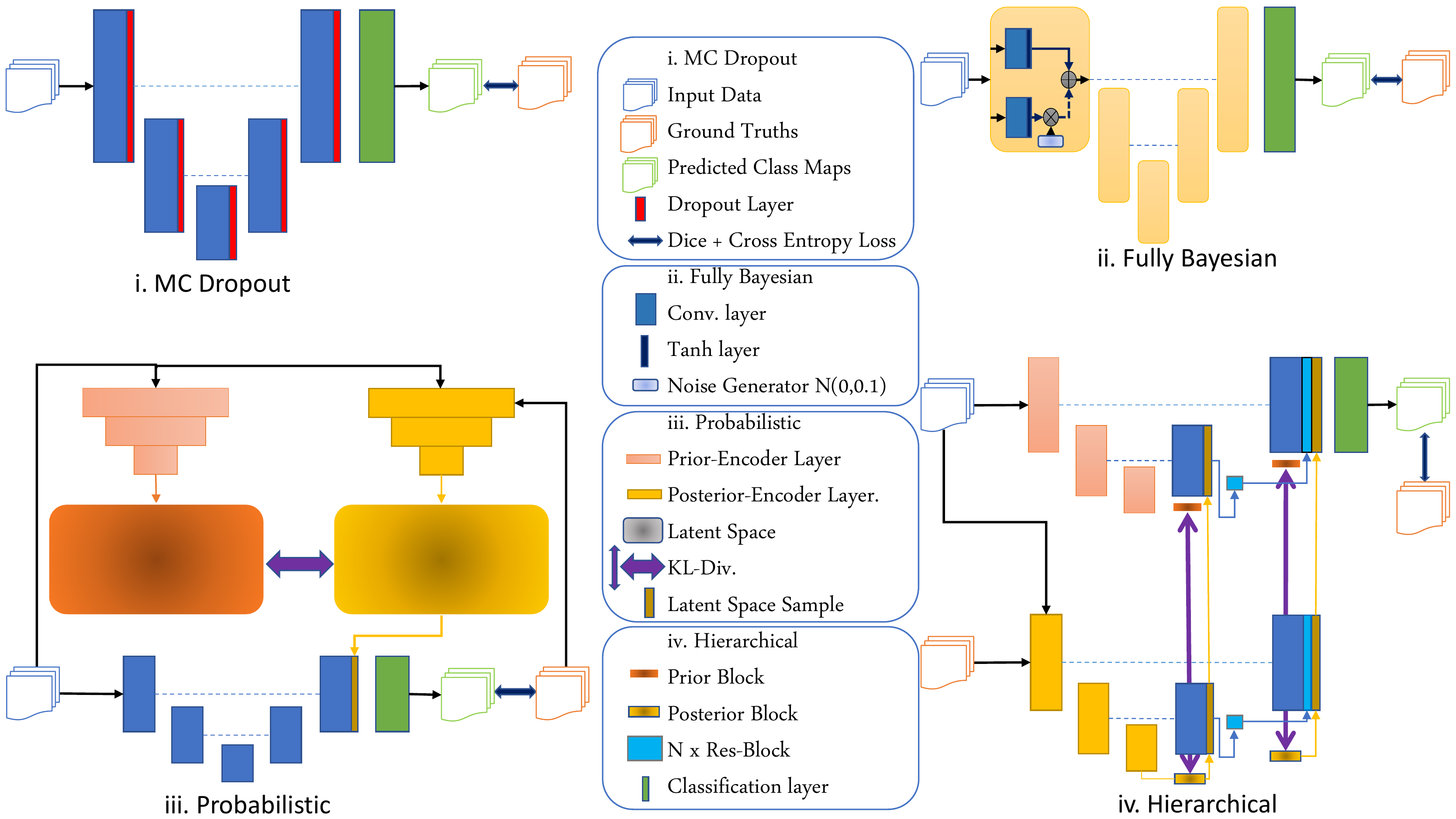}
\caption{Schematic illustration of the four Bayesian neural networks that we use for segmentation.\label{fig1:base_architecture}} 
\end{figure}

\subsubsection{Monte-Carlo Dropout}

Dropout layers were first adapted within deep neural networks to perform variational inference by Gal et al.~\cite{gal2016dropout}. This was later adopted for image segmentation in computer vision~\cite{kendall2017uncertainties} and medical imaging~\cite{roy2019bayesian}. The main idea is to keep the dropout layers active during inference, see Fig.~\ref{fig1:base_architecture}(i). This enforces the neurons to be stochastic within the network and generates multiple segmentation maps for a single image. These multiple Monte-Carlo segmentation samples are aggregated to generate the final segmentation and its corresponding uncertainty map.

\subsubsection{Fully-Bayesian F-CNN with Re-Parameterization}

We replace the convolutional layers in the segmentation network with a Bayesian convolution layer, see Fig.~\ref{fig1:base_architecture}(ii), which has been developed using the re-parameterization trick \cite{kingma2013auto, kingma2015variational}. 
We are not aware of a previous application of this approach for segmentation. 
The Bayesian layer consists of two convolution layers, whose outputs are further processed to add non-linearity by adding a tanh activation at each layer. 
We consider the outputs from the tanh layers as $\mu_\theta$ and $\sigma_\theta$. A Gaussian white noise $\varepsilon$  is multiplied with $\sigma_\theta$ to introduce stochasticity and the product is added to $\mu_\theta$ 
\begin{equation}
\label{rep_trick}
g_{\theta}(\varepsilon) = \mu_{\theta} + \varepsilon\sigma_{\theta}   \hspace{1cm}  \mbox{and} \hspace{1cm}     \varepsilon \sim \mathcal{N}(0, 0.1).
\end{equation}
Kingma et al.~\cite{kingma2013auto,kingma2015variational} used a Gaussian distribution with mean 0 and standard deviation 1 in their experiments. We reduce the standard deviation to 0.1 to  restrict higher variation in $\sigma_\theta$.

\subsubsection{Probabilistic U-Net}

Kohl et al.~\cite{kohl2018probabilistic} proposed the probabilistic U-Net. 
They suggested that ambiguous medical images can have multiple plausible segmentations based on multiple graders and that  capturing this spectrum of variability is more meaningful than estimating the uncertainty maps. 
In this regard, along with the base segmentation network, they train a separate network termed prior net, which maps the input to an embedding hypothesis space, see Fig.~\ref{fig1:base_architecture}(iii). One sample from this hypothesis space is concatenated to the last feature map of the base segmentation network to generate one segmentation output. Thus, multiple plausible segmentations are generated with sampling different points from the learnt hypothesis embedding space.

\subsubsection{Hierarchical Probabilistic U-Net}

Kohl et al.~\cite{kohl2019hierarchical} further improved the probabilistic U-Net~\cite{kohl2018probabilistic} to incorporate multi-scale ambiguity. In case the target organ  exists in multiple scales, it is important to capture the spectrum of variation across all the scales. The previous work from the authors successfully captured only the variation across one scale. Thus, they modified the network to capture the underlying variation across multiple scales. 
The main idea is to learn multiple hypothesis embedding spaces, each one specific to a specific target scale, see Fig.~\ref{fig1:base_architecture}(iv). 
In every encoder-decoder based F-CNN, it is assumed that different scale-specific features are learnt at each stage of the decoder with different spatial resolution. 
Therefore, during inference, multiple scale specific samples generated from different hypothesis embedding spaces are concatenated to their corresponding decoder feature map of appropriate scale. Consequently, different sets of samples from  different embeddings generate multiple plausible segmentation maps.

\subsection{Confidence Measure}

Bayesian neural networks commonly provide a measure of uncertainty per voxel. 
As many biomarkers are associated to organs, we need an uncertainty measure per organ. 
To this end, we will use the intersection over union (IoU) and the coefficient of variation (CV)~\cite{roy2019bayesian}. 
For the IoU, we consider $N$ segmentation samples $S_1, \ldots, S_N$ from the network for the organ $o$

\begin{equation}
\label{iou}
\mbox{IoU} = \frac{|(S_1 == o) \cap (S_2 == o) \cap ... (S_N == o)|}{|(S_1 == o) \cup (S_2 == o) \cup ... (S_N == o)|}.
\end{equation}
In our application, we use $N=10$ and $o = \mbox{liver}$.

In our analyses, we focus on the volume of the liver. 
Instead of quantifying uncertainty with regards to segmentation, we can also directly measure the variation of the volume across
the segmentation samples. 
Considering volumes $V_1, \ldots, V_N$ computed from the $N$ segmentation maps and the mean volume $\mu$, the  coefficient of variation is

\begin{equation}
\label{cv}
\mbox{CV} = \sqrt{\frac{\sum (V_i-\mu)^2}{N \cdot \mu^2}}.
\end{equation}

Note that this estimate is agnostic
to the size of the structure.
As a high coefficient of variation indicates an erroneous segmentation, we use the inverse, CV$^{-1}$, as confidence measure, while the IoU can be used directly. 

\subsection{Statistical Methods}

We want to integrate the segmentation confidence measures in the biomarker analysis to enable a more faithful and reliable estimation of model parameters. 
We present statistical models for group analysis and disease classification in the following. 

\subsubsection{Group Analysis}

In the group analysis, we evaluate the association of the biomarker with respect to non-imaging variables. For our application of diabetes, we consider the age $A_i$, sex $S_i$, BMI $B_i$, and diabetes status $D_i$ for subject $i$. The base model for the liver volume $V_i$ is

    \begin{equation}
    \label{r_eq0}
        \mbox{Base Model:} \quad V_i = \beta_0 + \beta_1  A_i + \beta_2  S_i + \beta_3  B_i + \beta_4  D_i + \varepsilon,
    \end{equation}
    where $\beta_0, \ldots, \beta_4$ are regression coefficients and $\varepsilon$ is the noise term. 
    
We now want to integrate the confidence measure $C_i$ that is associated to the volume $V_i$ and comes from the Bayesian segmentation into the model. 
As first approach, we propose to add the confidence measure as additional variable to the model

    \begin{equation}
    \label{r_eq1}
        \mbox{Variable:} \quad V_i = \beta_0 + \beta_1  A_i + \beta_2  S_i + \beta_3  B_i + \beta_4  D_i + \beta_5  C_i + \varepsilon.
    \end{equation}

As alternative, we use the confidence measure as instance weight in the regression model. Instead of giving equal importance to each subject~$i$, subjects with higher confidence in the liver segmentation will be given higher importance in the estimation of the coefficients

    \begin{equation}
    \label{r_eq2}
        \mbox{Instance Weighting:} \quad [V_i = \beta_0 + \beta_1  A_i + \beta_2  S_i + \beta_3  B_i + \beta_4  D_i + \varepsilon] \cdot C_i.
    \end{equation}
    
Weighted least squares is used for estimating the model coefficients, where the confidence measures are used as weights. 
Note that manual quality control is a special case of instance weighting with a binary confidence measure, where only those segmentations with passing quality are set to one and the rest to zero. 

\subsubsection{Disease Classification}

For the prediction of the diabetes status, we use logistic regression. 
In the base model, we use the liver volume as input feature. 
For the integration of the confidence measure in the classification model, we consider three variants. 
First, we add the confidence measure $C_i$ as additional variable to the model. 
Second, we do not only consider the additive effect of the confidence measure but also the interaction, so that $V_i \cdot C_i$ is also added to the model. 
Third, we use instance weighting based on the confidence measure to emphasize subjects with good segmentations.

\section{Results}
\subsection{Data}

Experiments are performed on a set of whole-body Magnetic Resonance Images (MRI) obtained from the Cooperative Health Research in the Region Augsburg project (KORA). 
We work with 308 subjects (109 diabetic, 199 non-diabetic)  that have a manual annotation of the liver. 
We resample all volumes to a standard resolution of $2 \times 2 \times 3$ mm\textsuperscript{3} and $53 \times 256 \times 144$ voxels.

\begin{figure}[t]
    \centering
    \subfloat[Liver Dice Score]{\includegraphics[height=4cm]{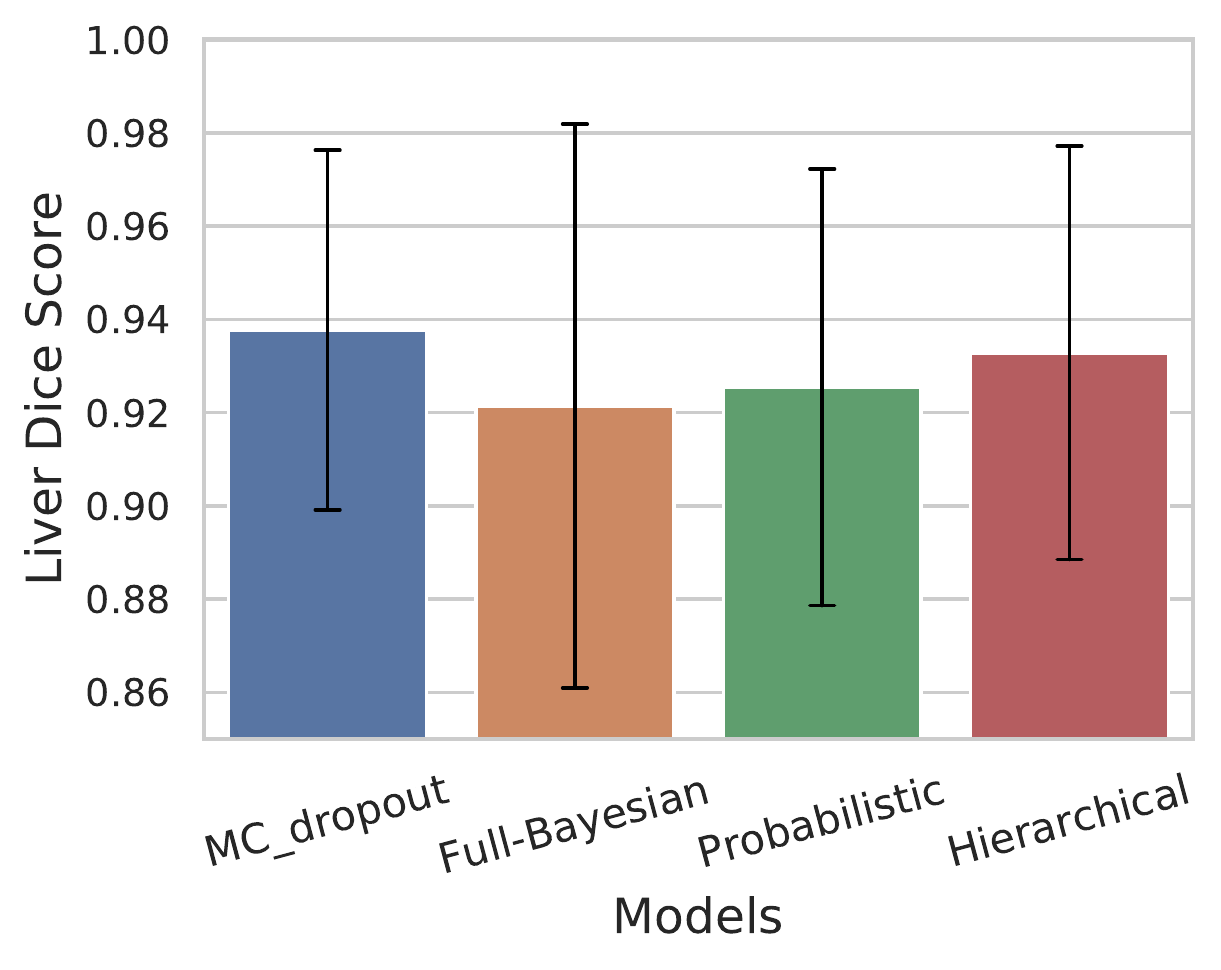} }%
    \qquad
    \subfloat[Liver Volume per Group]{\includegraphics[height=4cm]{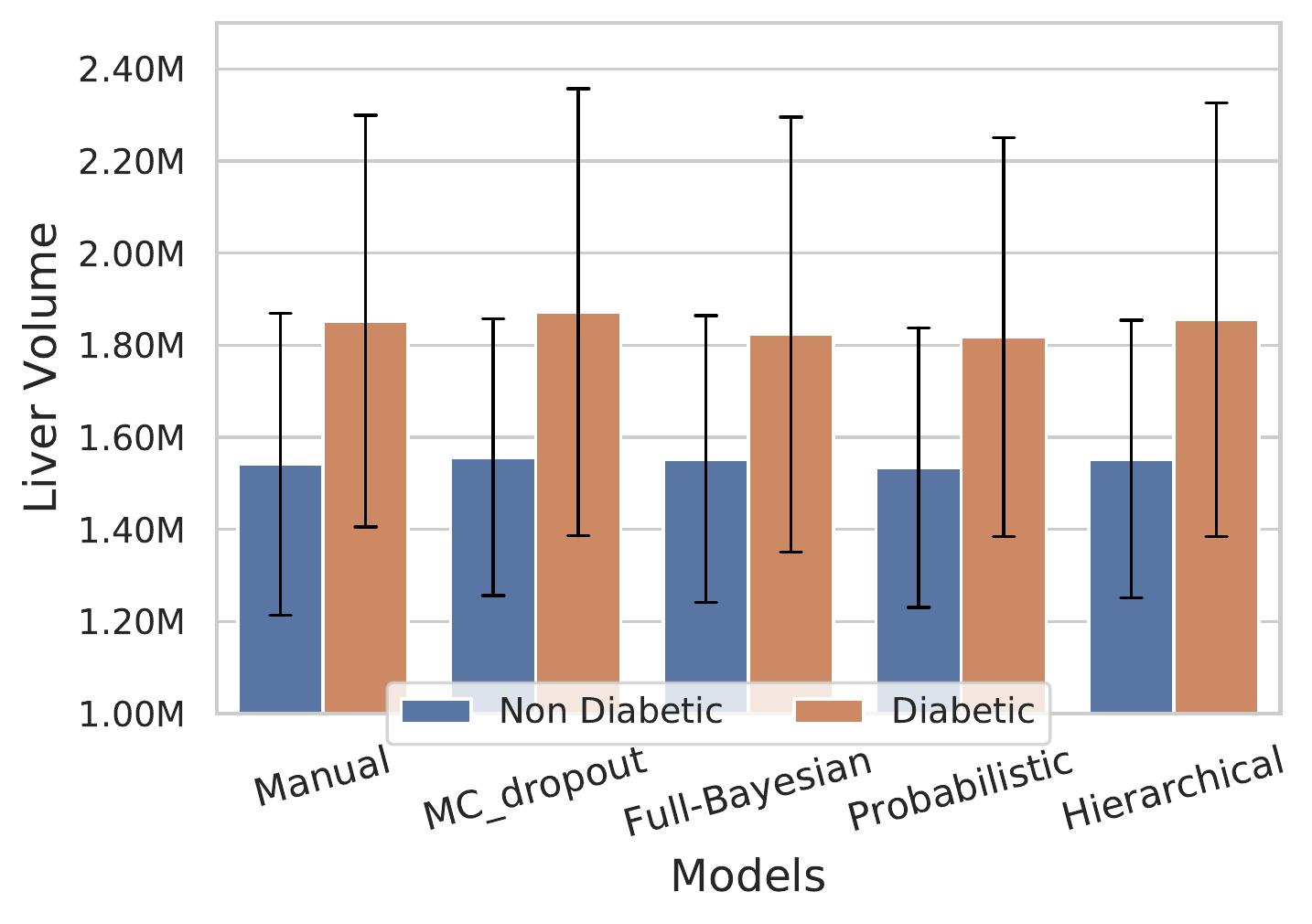}}%
    \vspace{-2mm}
    \caption{(a) Bars show mean Dice score of liver and error bars show standard deviation for different segmentation networks. (b) Bars show the mean liver volume of diabetic and non-diabetic subjects for manual and automated segmentation, error bars indicate standard deviation.\vspace{-2mm}}%
    \label{fig:dice_and_volume}%
\end{figure}

\subsection{Bayesian segmentation results}

We use QuickNAT~\cite{roy2017error,roy2019quicknat} as base architecture for implementing Bayesian  networks in Fig.~\ref{fig1:base_architecture}. 
We use common network parameters across  models with learning rate 1e-5, batch size 5, and 50 epochs.
For MC Dropout, we use a dropout rate of 0.2. 
For Fully-Bayesian, we do not use a batch normalization layer. Instead, we use uni-variate KL-Divergence loss to regularize the distribution of weights from each Bayesian layer. 
For the probabilistic and hierarchical models, a latent variable of dimension 12 has been used to estimate the posterior embedding.
We split the dataset into 155 training (56 diabetic, 99 non-diabetic) and 153 testing (53 diabetic, 100 non-diabetic) subjects by equally distributing diabetic and non-diabetic subjects. 

\begin{figure}[t]
\includegraphics[clip, trim=0cm 5cm 0cm 0cm, width=1.00\textwidth]{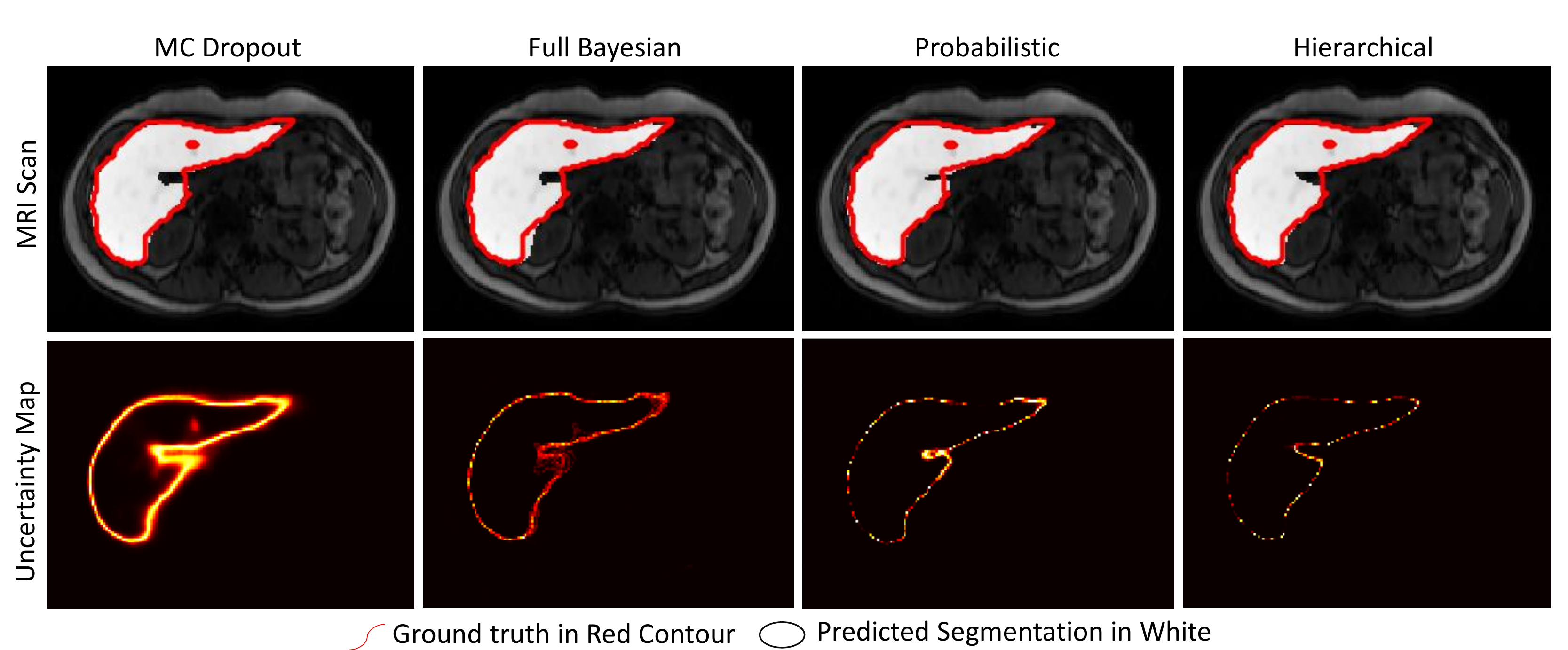}
\caption{Top: MRI scan overlaid by segmentation map (white) and manual annotation (red contour). 
Bottom: Voxel-wise uncertainty map of the segmentation.\vspace{-3mm}} 
\label{fig1:uncertainty_maps}
\end{figure}

Fig.~\ref{fig:dice_and_volume}(a) shows a boxplot of the Dice score of the liver for different segmentation methods. We observe that MC dropout yields the highest accuracy, followed by hierarchical, probabilistic, and Fully-Bayesian. 
Overall, the performance of the models is high, but the error bars indicate that accuracy substantially varies across subjects.
Fig.~\ref{fig1:uncertainty_maps} visualizes the predictions and uncertainty maps for the different segmentation methods. We observe that MC Dropout and Fully-Bayesian give  a higher uncertainty in comparison to the other two. 
The probabilistic and hierarchical models were designed to learn annotations from multiple raters, while we only have annotations from a single rater, which may explain the lower stochasticity of these models in our experiments. 

\subsection{Group analysis}

Table~\ref{regression_tbl} reports the regression coefficient of the diabetes status $\beta_4$, which is of primary interest for studying diabetes, where Fig.~\ref{fig:dice_and_volume}(b) illustrates that diabetic subjects tend to have a higher liver volume. 
The table shows the coefficient for different segmentation methods and statistical models. 
Next to the base model, and the integration of the confidence measure as additional variable and instance weight, we also estimate the coefficient from the manual segmentation, which serves as reference.  
All regression coefficients are estimated on the segmentation test set. 

MC Dropout with IoU as additional variable or instance weight yields coefficients that are closest to the manual estimate. 
For the probabilistic and hierarchical models, adding IoU to the model leads to the best results. 
The coefficient of the Fully-Bayesian network has the highest divergence from the manual estimate across segmentation methods, but inclusion of confidence measures as variables helps to improve the estimate. 

\begin{table}
\centering
\caption{Regression coefficient of diabetes status, $\beta_4$, for different segmentation approaches and statistical models, together with the manual estimate.  \label{regression_tbl}}
\begin{tabular}{@{} l
                @{\hspace*{\lengtha}}l
                @{\hspace*{\lengtha}}c
                @{\hspace*{\lengtha}}c
                @{\hspace*{\lengtha}}c
                @{\hspace*{\lengtha}}c
                @{\hspace*{\lengtha}}c @{}}  \toprule
  & \multirow{2}{*}{Base} & \multicolumn{2}{c}{Variable} & \multicolumn{2}{c}{Instance} & \\ 
    \cmidrule(r){3-4} \cmidrule(r){5-6}
                 &       & IoU & CV\textsuperscript{-1} & IoU & CV\textsuperscript{-1} & Manual \\  \midrule
 MC Dropout      & 0.308 & {\bfseries 0.318} & 0.308             & 0.316 & 0.268 & \multirow{4}{*}{0.328} \\ 
 Fully-Bayesian  & 0.255 & 0.269             & {\bfseries 0.271} & 0.254 & 0.229 &  \\
 Probabilistic   & 0.287 & {\bfseries 0.302} & 0.297             & 0.287 & 0.192 & \\ 
 Hierarchical    & 0.294 & {\bfseries 0.306} & 0.288             & 0.295 & 0.249 & \\ \bottomrule
\end{tabular}
\end{table}

\subsection{Disease classification}

For the classification experiment, we split the segmentation test set further into a classification training set (77 subjects, 27 diabetic, 50 non-diabetic) and a classification test set (76 subjects, 26 diabetic, 50 non-diabetic) randomly 1,000 times. 
Table \ref{classification_tbl} reports the mean classification accuracy across all runs. 
We compare different methods for segmentation  and  integration of the confidence measures. 

The accuracy for volumes derived from manual annotations is 0.713; the accuracy decreases for all automated segmentations in the base model. 
The inclusion of the confidence measures in the classification helps to recover the accuracy and even pass the one from the manual segmentation. 
The likely reason for this behaviour is that the confidence measures vary between diagnostic groups and therefore provide additional information for the classification. 
We observe the best performance for the variable and interaction models with IoU. 
The overall best result is obtained by the probabilistic model with IoU interaction term. 

\begin{table}[t]
\centering
\caption{Accuracy for diabetes classification with logistic regression for different segmentation methods and manual segmentation. The base model is compared to several approaches of including the confidence measure in the estimation.  \label{classification_tbl}}
\begin{tabular}{@{} l
                @{\hspace*{\lengtha}}l
                @{\hspace*{\lengtha}}c
                @{\hspace*{\lengtha}}c
                @{\hspace*{\lengtha}}c
                @{\hspace*{\lengtha}}c
                @{\hspace*{\lengtha}}c
                @{\hspace*{\lengtha}}c
                @{\hspace*{\lengtha}}c
                @{\hspace*{\lengtha}}c@{}} \toprule
  & \multirow{2}{*}{Base} & \multicolumn{2}{c}{Variable} & \multicolumn{2}{c}{Interaction} & \multicolumn{2}{c}{Instance} &  \multirow{2}{*}{Manual} \\ 
  \cmidrule(r){3-4} \cmidrule(r){5-6}\cmidrule(r){7-8}
   &  & IoU & CV\textsuperscript{-1} & IoU & CV\textsuperscript{-1} & IoU & CV\textsuperscript{-1}\\  \midrule
 MC Dropout     & 0.702 & {\bfseries 0.719} & 0.709 & 0.716  & 0.712 & 0.706 & 0.708 & \multirow{4}{*}{0.713} \\ 
 Fully-Bayesian & 0.692 & {\bfseries 0.705} & 0.696 & {\bfseries 0.705} & 0.695 & 0.695 & 0.696  \\
 Probabilistic  & 0.691 &            0.719  & 0.696 & {\bfseries 0.732} & 0.694 & 0.691 & 0.696 \\ 
 Hierarchical   & 0.702 & {\bfseries 0.714} & 0.694 & {\bfseries 0.714} & 0.695 & 0.703 & 0.699  \\ \bottomrule
\end{tabular}
\end{table}

\section{Conclusion}

In this work, we proposed to propagate segmentation uncertainty to the bio-marker analysis as the segmentation quality can vary substantially between scans. Our results have demonstrated that assigning a confidence score to an imaging biomarker can yield a more faithful estimation of model parameters and a higher classification accuracy.
We have evaluated four Bayesian neural networks with the best results for MC dropout and the probabilistic model, each one in combination with IoU as confidence measure. 
These results show a clear improvement over a base model that does not consider segmentation uncertainty, and therefore confirms the necessity of propagating uncertainty to the final biomarker analysis.

\noindent
\textbf{Acknowledgement:} This research was supported by DFG, BMBF (project DeepMentia), and the Bavarian State Ministry of Science and the Arts and coordinated by the Bavarian Research Institute for Digital Transformation (bidt).

\end{document}